# Role of an additional interfacial spin-transfer torque for current-driven skyrmion dynamics in chiral magnetic layers

Callum Robert MacKinnon[1], Serban Lepadatu[2], Tim Mercer[3], Phillip Bissell[4]

*Jeremiah Horrocks Institute for Mathematics, Physics and Astronomy, University of Central Lancashire, Preston PR1 2HE, U.K.*

Abstract

Skyrmions can be driven by spin-orbit torques as a result of the spin Hall effect. Here we model an additional contribution in ultra-thin multilayers, arising from the spin accumulation at heavy metal / ferromagnetic interfaces and observe the effects on a large range of skyrmion diameters. The combination of the interfacial spin-transfer torque and the spin-orbit torque results in skyrmion motion which helps to explain the observation of small skyrmion Hall angles for skyrmion diameters less than 100 nm. We show that this additional term has a significant effect on the skyrmion dynamics and leads to rapidly decreasing skyrmion Hall angles for small skyrmion diameters, as well as a skyrmion Hall angle versus skyrmion velocity dependence nearly independent of the surface roughness characteristics. Also, the effect of various disordered energy landscapes, in the form of surface roughness, on the skyrmion Hall angle and velocity is shown to be largely drive-dependent. Our results show good agreement with those found in experiments thus concluding that the interfacial spin-transfer torque should be included in micromagnetics simulations for the reproduction of experimental results.

[1] CRMackinnon@uclan.ac.uk
[2] SLepadatu@uclan.ac.uk
[3] TMercer1@uclan.ac.uk
[4] PBissell@uclan.ac.uk

# I. Introduction

Skyrmions are topologically protected quasi-particle-like magnetic structures [1], which are of great interest as they can be efficiently manipulated using spin torques due to applied currents in ultra-thin magnetic multilayers, thus opening the possibility of a new class of spintronics devices [2-4]. Skyrmions have been observed at room temperature in systems with broken inversion symmetry [5-9], stabilized by the interfacial Dzyaloshinskii-Moriya interaction (DMI) [10,11], and more recently current-induced skyrmion motion has been demonstrated in ultra-thin magnetic multilayers [12-19] by utilizing spin currents due to the spin-Hall effect (SHE) which gives rise to spin-orbit torques (SOT) [20-24]. The SHE governs SOT via the accumulation of spins at the boundaries of a current-carrying non-magnetic heavy metal (NM) / ferromagnetic (FM) bilayer due to the generation of transverse spin currents. The SHE exploits the spin-orbit coupling in a NM to convert the unpolarized charge into a pure spin current which arises from both intrinsic and extrinsic effects and gives rise to an asymmetric scattering of conduction electrons. Irrespective of the underlying origins of the SHE and thus spin accumulation at the NM/FM interface, the SOT exerted on the FM magnetization can be broken down into two components, namely the damping-like (DL) SOT and the field-like (FL) SOT.

Current-driven skyrmion motion is currently hindered by the skyrmion Hall angle (SkHA), $\theta_{SkHA}$, in which the skyrmion path deviates from the direction of the applied charge current, which poses problems for spintronics applications. Furthermore, the SkHA has been shown to vary with the applied current strength [15-19], and this effect is not reproduced by analytical or numerical models in ideal structures. Whilst material imperfections play a very important role [25], there are many gaps in the literature, and a full understanding of how the SkHA varies with driving strength, skyrmion diameter, skyrmion chirality, and varying levels of magnetic and non-magnetic imperfections is required. In recent experimental work, a largely independent SkHA on skyrmion diameter has been observed [14] however analytical results based on the Thiele equation predict a significant increase in SkHA with decreasing skyrmion diameter, and this discrepancy is significantly pronounced for small diameters below 100 nm. In another study [18] experimental results were also compared to modelling based on SOT for very large skyrmions with diameters above 800 nm. Experimental results and SOT simulations show good agreement for large skyrmions diameters, which showed slightly larger SkHA values for the SOT model, however, the SOT model predictions deviate significantly from the experimental results at small skyrmion diameters [14] even when very large damping values are assumed. Reducing the SkHA is critical for spintronic devices and it has been theoretically shown that the spin-orbit coupling in a NM/FM bilayer can be used to tune the skyrmion Hall effect by controlling gate voltages to steer skyrmions [26], and furthermore adjusting the strength of the Rashba spin-orbit coupling has been show to result in a negligible SkHA [27].

Very recently, it has been proposed that interfacial spin-transfer torques (ISTT) [28] may also be important, and thus a more complete model based on both SOT and ISTT can be formed. Here we show, using micromagnetics modelling coupled with a self-consistent spin transport solver in multilayers, the addition of ISTT results in a significantly reduced discrepancy between experimental and modelling results, and a good agreement is obtained with recent experimental results which have demonstrated a nearly diameter-independent SkHA [14]. In this paper, we study the effect of the additional ISTT term on skyrmion motion in a Co (1 nm)/Pt (3 nm) bilayer with and without surface disorder in the form of surface roughness as a function of skyrmion diameter. An overview of the spin transport model is given in Section II. In Section III we study the disorder-free case, investigating skyrmion motion as a function of skyrmion diameter and interfacial spin mixing conductance. Finally, in Section IV we also consider the effect of surface roughness on skyrmion motion, where the SkHA is shown to vary with current density as the skyrmion motion changes from pinned, to creep, and finally plastic flow regime. The results obtained for a range of skyrmion diameters are shown to collapse onto a universal SkHA versus skyrmion velocity dependence, nearly independent of the surface roughness characteristics.

## II. Spin Transport Model

The effect of spin torques on skyrmion motion is modelled using the Landau-Lifshitz-Gilbert (LLG) equation:

$$\frac{\partial \mathbf{m}}{\partial t} = -\gamma \mathbf{m} \times \mathbf{H}_{eff} + \alpha \mathbf{m} \times \frac{\partial \mathbf{m}}{\partial t} + \frac{1}{M_S} \mathbf{T_S} \tag{1}$$

Here $\gamma = \mu_0 g_{rel} |\gamma_e|$, where $\gamma_e$ is the electron gyromagnetic ratio, $g_{rel}$ is a relative gyromagnetic factor, $\alpha$ is the Gilbert damping, and $M_s$ is the saturation magnetization. $\mathbf{H}_{eff}$ is the effective field term consisting of several additive terms, namely the demagnetizing field, exchange interaction including the interfacial Dzyaloshinskii-Moriya exchange interaction in the xy plane, applied magnetic field, and uniaxial magnetocrystalline anisotropy. $\mathbf{T}_s$ is the total spin-torque term, which includes contributions from SOT and ISTT, and can be computed self-consistently using the drift-diffusion model where the spin current density is given as:

$$\mathbf{J}_S = -\frac{\mu_B}{e} P \sigma \mathbf{E} \otimes \mathbf{m} - D_e \nabla \mathbf{S} + \theta_{SHA} \frac{\mu_B}{e} \boldsymbol{\varepsilon} \sigma \mathbf{E} \tag{2}$$

Here, the drift-diffusion model shows three distinct components, where $\mathbf{J}_s$ is the spin current such that $\mathbf{J}_{Sij}$ indicates the flow of the j component of spin polarization in the i direction. The first is the contribution due to drift included in the FM layers, where $P$ is the current spin-polarization, $\sigma$ is the electrical conductivity, $\mathbf{m}$ is the normalized magnetization and $\mathbf{E}$ is the electric field. The second is a diffusive term, where $D_e$ is the electron diffusion constant and $\mathbf{S}$ is the spin accumulation. The third term is the SHE in the NM layers, where $\theta_{SHA}$ is the spin-Hall angle and $\boldsymbol{\varepsilon}$ is the rank-3-unit asymmetric tensor. The drift-diffusion model also includes the charge current density, given as $\mathbf{J}_C = \sigma \mathbf{E}$, which is simplified here by assuming the inverse SHE is negligible.

Transverse components of spin currents are absorbed at a NM/FM interface resulting in a spin-torque which can be included in the drift-diffusion model with the incorporation of the spin mixing conductance, $G^{\uparrow\downarrow}$, with the resulting total spin-torque obtained using circuit theory boundary conditions [29] as:

$$\mathbf{T}_S = \frac{g\mu_B}{ed_F}\left[\text{Re}\{G^{\uparrow\downarrow}\}\mathbf{m}\times(\mathbf{m}\times\Delta\mathbf{V}_S) + \text{Im}\{G^{\uparrow\downarrow}\}\mathbf{m}\times\Delta\mathbf{V}_S\right] \quad (3)$$

Here $d_F$ is the ferromagnetic layer thickness, $\Delta\mathbf{V}_S$ is the spin chemical potential drop across a NM/FM interface, where $\mathbf{V}_S = (D_e/\sigma)(e/\mu_B)\mathbf{S}$. The drift-diffusion model can be solved analytically to obtain the following expression for the SOT, given in Equation (4), by assuming negligible in-plane spin diffusion.

$$\mathbf{T}_{SOT} = \theta_{SHAeff}\frac{\mu_B}{e}\frac{|J_c|}{d_F}[\mathbf{m}\times(\mathbf{m}\times\mathbf{p}) + r_G\mathbf{m}\times\mathbf{p}] \quad (4)$$

The SOT contains both damping-like and field-like components, with $r_G$ being the FL-SOT coefficient. Here $\mathbf{p} = \mathbf{z}\times\mathbf{e}_{Jc}$, where $\mathbf{e}_{Jc}$ is the charge current direction. The quantity $\theta_{SHAeff}$ is the effective spin-Hall angle, proportional to the real, or intrinsic, spin-Hall angle, $\theta_{SHA}$; $\theta_{SHAeff}$ and $r_G$ are given in Equations (5) and (6) respectively.

$$\theta_{SHAeff} = \theta_{SHA}\left(1 - \frac{1}{\cosh(d_N/\lambda_{sf}^N)}\right)\frac{\text{Re}\{\tilde{G}\}^2 - \text{Im}\{\tilde{G}\}^2 + N_\lambda \text{Re}\{\tilde{G}\}}{(N_\lambda + \text{Re}\{\tilde{G}\})^2 + \text{Im}\{\tilde{G}\}^2}, \quad (5)$$

$$r_G = \frac{N_\lambda \text{Im}\{\tilde{G}\} + 2\text{Re}\{\tilde{G}\}\text{Im}\{\tilde{G}\}}{N_\lambda \text{Re}\{\tilde{G}\} + \text{Re}\{\tilde{G}\}^2 - \text{Im}\{\tilde{G}\}^2} \quad (6)$$

Here $N_\lambda = \tanh(d_N/\lambda_{sf}^N)/\lambda_{sf}^N$ and $\tilde{G} = 2G^{\uparrow\downarrow}/\sigma_N$. As shown previously [30], it is important to make the distinction between the intrinsic spin Hall angle, and the effective spin Hall angle at the NM/FM interface, as the latter can be significantly smaller depending on the interface transparency as modelled using the spin mixing conductance.

Another important source of vertical spin current is due to the NM/FM inter-layer diffusion of a spin accumulation generated in the FM layer, resulting in an interfacial spin-torque contribution. In ultra-thin films the inter-layer spin diffusion results in spin torques similar in form to the bulk Zhang-Li STT [31,32], but greatly enhanced partly due to the inverse dependence on $d_F$ and acting in the opposite direction. This additional interfacial spin torque is given in Equation (7).

$$\mathbf{T}_{iSTT} = -\left[(\mathbf{u}_\perp \cdot \nabla)\mathbf{M} - \frac{\beta_\perp}{M_S}\mathbf{M} \times (\mathbf{u}_\perp \cdot \nabla)\mathbf{M}\right] \quad (7)$$

The interfacial spin-torque acts in the opposite direction to the Zhang-Li STT, and the spin-drift velocity and the non-adiabaticity parameter are replaced by an effective perpendicular spin-drift velocity, Equation (8), and effective non-adiabaticity parameter.

$$\mathbf{u}_\perp = \mathbf{J}_C \frac{|P_\perp| g\mu_B}{2eM_S} \frac{1}{1+\beta_\perp^2} \quad (8)$$

Here $P_\perp$ is an effective perpendicular spin polarization parameter, which in the case of ISTT takes on negative values. These values are not dependent on a single material alone but depend on the transport properties of both the NM and FM layers [28].

The ISTT arises due to an imbalance in the spin accumulation either side of the NM/FM interface. This imbalance gives rise to diffusive vertical spin currents, which contain both longitudinal and transverse spin components. The transverse spin components are absorbed at the interface, and due to conservation of total angular momentum this results in an interfacial spin torque. The ISTT thus arises due to absorption of transverse spin components, as does the SOT, but in this case the vertical spin currents originate in the FM layer, not the NM layer. The spin imbalance arises due to the spin accumulation generated at magnetization gradients and is due to the non-zero divergence of the in-plane spin current in the FM layer, as well as the spin mistracking effect due to the exchange rotation and spin dephasing processes. This is the same transverse spin accumulation which gives rise to the Zhang-Li adiabatic and non-adiabatic spin transfer torques, however the ISTT arises due to absorption of vertical, transverse spin components at the interface, not due to the in-plane, transverse spin absorption lengths, and is thus an interfacial spin torque with strength inversely dependent on the FM layer thickness. We note that experimental evidence for a huge negative spin transfer torque was obtained for ultrathin Co layers interfaced with Pt [33]. This torque was found to act along the current direction with a negative spin polarization, scaling inversely to the Co layer thickness, and for ultrathin Co layers its strength far exceeds that of the bulk STT and is comparable to or greater than the SOT, thus similar to the ISTT torque in Equation (7) discussed here. Our results could provide an explanation for these observations, however the analysis of this case is left for future work.

Experimental results are often compared to an analytical model known as the Thiele equation [34], which can be used to describe the dynamics of magnetic skyrmions due to various driving forces. There is an included caveat in assuming skyrmions have rigid body magnetic characteristics, however, this allows the skyrmions to be considered with steady-state velocity arising from the equilibrium between different forces acting on the rigid body. The Thiele equation includes (i) the force acting on the skyrmion; in the bilayer systems, this total force results from the DL-SOT and ISTT. (ii) The skyrmion Hall effect, whereby the skyrmion motion deviates from the charge current direction due to the Magnus force [35]. (iii) The dissipative force, which includes a dissipative tensor and the Gilbert damping coefficient. The solution to the Thiele equation is shown in Equation (9), which gives an expression for the diameter-dependent and drive-independent SkHA [15,36].

$$v = \sqrt{\frac{(u_{SOT} - \beta D u_{STT})^2 + u_{STT}^2}{1 + (\alpha D)^2}}$$

$$\tan(\theta_{SkHA}) = \frac{u_{SOT} + (\alpha - \beta) D u_{STT}}{\alpha D u_{SOT} - (1 + \alpha \beta D^2) u_{STT}}$$

(9)

Here $D = R/2\Delta$, where $\Delta$ is the domain wall width and $R$ is the skyrmion radius. Moreover $u_{SOT} = \mu_B \theta_{SHAeff} \pi R J_{NM} / 4 e M_s d_F$, and $u_{STT} = P g \mu_B J_{FM} / 2 e M_s (1+\beta^2)$, where $J_{NM}$ and $J_{FM}$ are the current densities in the NM and FM layers respectively. Equation (9) models the contribution of both the SOT and ISTT for current-driven skyrmion motion and is in agreement with micromagnetics modelling for the ideal skyrmion case as we have verified.

## III. Spin Torques in a Disorder-Free System

Current-induced Néel skyrmion movement has been observed in asymmetric stacks including [Pt/Co/Ta]$_x$ [12], Pt/GdFeCo/MgO [13], Ta/CoFeB/TaO$_x$ [37], Pt/Co/Ir [17] as well as more complex stacks such as Ta/Pt/[Pt/CoB/Ir]$_x$Pt [14] and symmetric stacks [38]. To study the effect of the SOT and ISTT on skyrmion motion, a Pt(3 nm)/Co(1 nm) bilayer thin-film [39,40] structure was chosen for this work, using periodic boundary conditions for the demagnetizing field. The current was applied across the bilayer through electrodes attached on its x-axis ends.

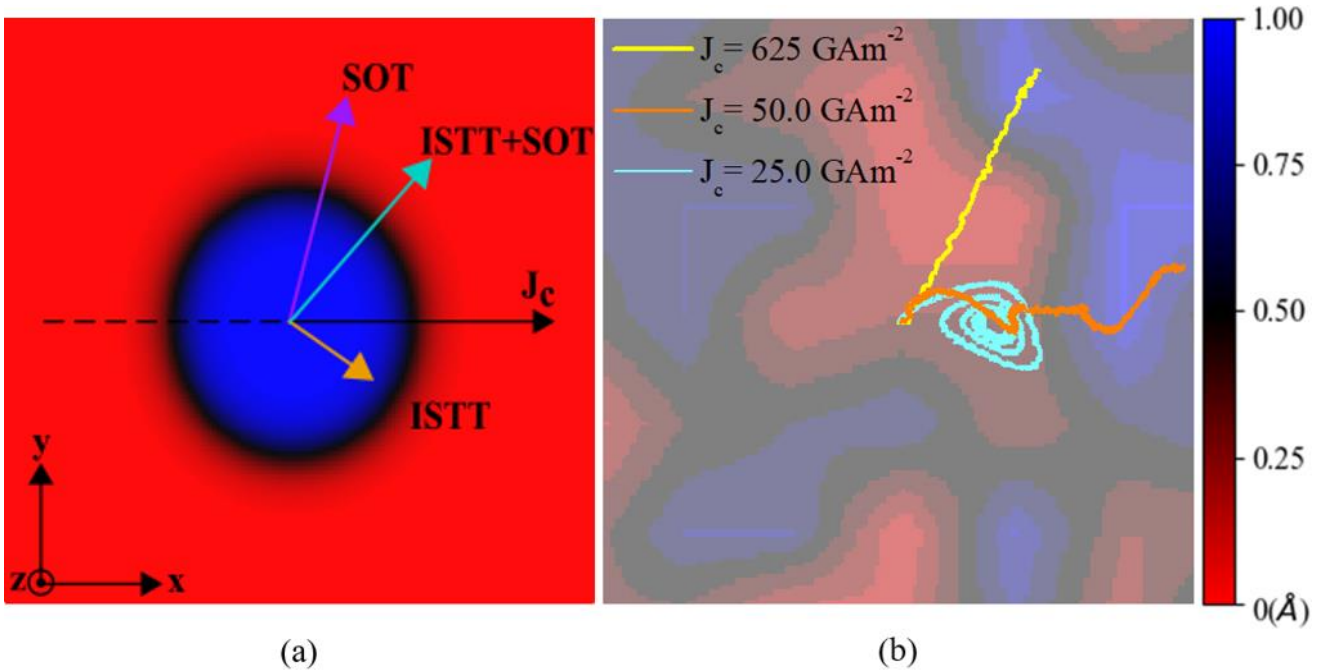

FIG. 1. (a) A snapshot of a 400 nm diameter Néel skyrmion, showing the z magnetization component. The magenta, orange and cyan arrows represent the direction in which the SOT, ISTT and the combination of these two torques in the spin transport solver act on the skyrmion when a current is applied, respectively. The black arrow indicates the conventional current direction, $\mathbf{J_c}$. (b) Skyrmion motion in a bilayer system displaying pinned, creep and flow regimes in cyan, orange, and yellow, respectively. The contoured effect outlines the disordered pattern on the surface of the sample with depth displayed by the heatmap. All paths were obtained using a damping constant $\alpha = 0.03$ and disorder variation period of 60 nm.

Fig. 1(a) illustrates the SOT and ISTT components acting on a current-driven skyrmion with the combined effect displayed, which is a vector addition of the two torques. To investigate the effect of spin torques on an isolated skyrmion, first, a disorder-free bilayer was used, neglecting thermal fluctuations and material imperfections, with a fixed current density. The skyrmion diameter is controlled using an out-of-plane magnetic field in the +z direction ranging from $|H_z| = 0.875$ kAm$^{-1}$ to $|H_z| = 54$ kAm$^{-1}$, which yielded diameters ranging from 396 nm to 21 nm, respectively. The full dependence of skyrmion diameter on the applied field is shown in Appendix A. The results for the ideal case are shown in Fig. 2. For the spin torques obtained with the self-consistent spin transport solver, we show two distinct contributions and the combined torque effect. The contributions observed are the SOT due to SHE, as well as another important contribution obtained due to the inter-layer diffusion of spin currents, namely the ISTT. To obtain solutions for these contributions, $P_\perp$, $\beta_\perp$, $r_G$, and $\theta_{SHAeff}$ were obtained numerically. Thus the spin accumulation was computed using the spin transport solver, and the total spin-torque obtained from Equation (3). Then, Equation (4) and Equation (7) for the SOT and ISTT respectively, were used to obtain the above constants as fitting parameters to the total spin torque. Whilst $\theta_{SHAeff}$ and $r_G$ do not depend on the skyrmion diameter, $P_\perp$ and $\beta_\perp$ are not constant and show a small dependence on skyrmion diameter due to spin diffusion, similar to the enhancement of Zhang-Li non-adiabaticity obtained for vortex cores [41,42]. The dependence of these parameters on skyrmion diameter is shown in Appendix B. Fig. 2(a) shows three clear distinct curves which are identified by the dashed, solid, and dash-dot lines which represent the SOT, the full spin torque and ISTT, respectively. There is a clear difference in each line suggesting each component has a significant contribution to the skyrmion motion. The ISTT has a significant effect on the SkHA which is only applicable for ultrathin films due to the inverse thickness dependence, and NM layers with a small spin diffusion length, such as the Co/Pt bilayer investigated here. Fig. 2(b) shows the skyrmion path for each component, which shows that the SOT has the largest effect on the skyrmion motion compared to the ISTT component, as can be deduced from the SOT and ISTT skyrmion displacements. For large diameters the SOT and full torque dependencies tend towards each other and become near identical. This is an important result as it shows the SOT-modelled results for the skyrmion motion to be a good assumption for large diameters. However, for small skyrmion diameters, the two trends deviate rapidly, which would explain discrepancies between experimental and theoretical results. The combined torque results correspond well with results obtained experimentally both for small skyrmion diameters [14] as well as for large diameters [14,15,18].

We note the ISTT is not confined to Néel type skyrmions alone as the ISTT arises from spin accumulation at magnetization gradients and is therefore applicable for any magnetic texture including anti-skyrmions, skyrmioniums, and domain walls, however investigation of these cases is outside the scope of this work.

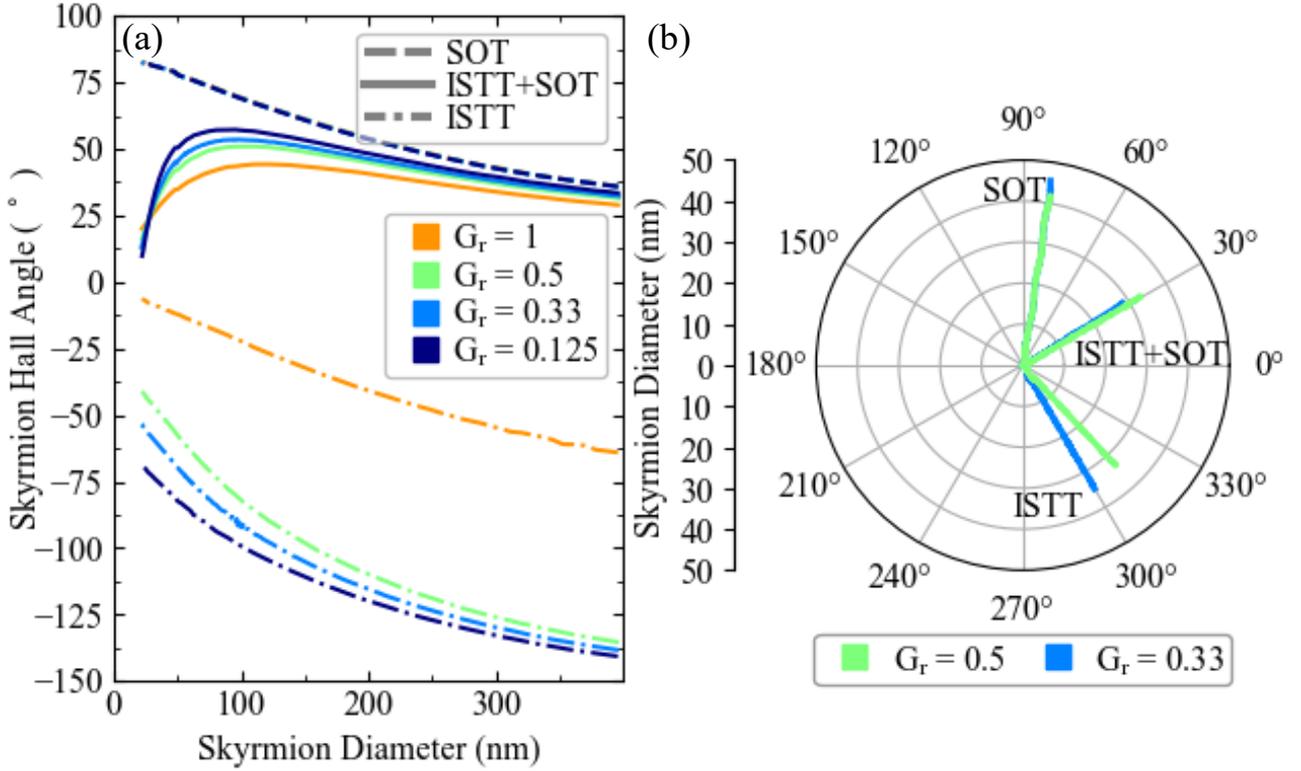

FIG. 2. Relationship between the skyrmion Hall angle and skyrmion diameter for different imaginary to real spin mixing conductance ratios, $G_r$. (a) Individual SOT and ISTT contributions for each ratio represented by a dashed and dash-dot line, respectively, as well as the full spin transport solver solution including both SOT and ISTT shown by the solid lines. (b) Skyrmion paths for $G_r = 0.33$ and $0.5$. The individual SOT and ISTT contributions are shown alongside the full spin transport solver path. The current is applied along the x-direction, 0º. To obtain these results a damping value of $\alpha = 0.1$ was used.

The imaginary part of the spin mixing conductance, $\text{Im}(G^{\uparrow\downarrow})$, is commonly neglected with an effective spin mixing conductance used instead. The imaginary component is usually assumed to be much smaller than the real component, however imaginary to real spin mixing conductance ratios, $G_r = \text{Im}(G^{\uparrow\downarrow})/\text{Re}(G^{\uparrow\downarrow})$, larger than one have been measured recently [43]. Fig. 2(a) shows the dependence of SkHA on the spin mixing conductance ratio with a clear influence from the ratio apparent. The SOT is not affected by $G_r$ in terms of the SkHA, but skyrmion velocity is affected by the $G_r$ ratio, as is apparent in Fig. 2(b) in which the velocity is larger for $G_r = 0.33$. Furthermore, the $G_r$ ratio affects the ISTT contribution of both velocity and SkHA. The peaks of the combined torque curves range from 44° to 57.3° for $G_r = 1$ and $G_r = 0.125$, respectively. Fig. 2(b) shows the skyrmion paths for each component for $G_r = 0.33$ and 0.5, illustrating further the velocity dependence on $G_r$. The ratio used in all subsequent modelling throughout this paper is determined using both DL- and FL-SOT components, values of which are given in Ref. [19] for a Pt/Co (0.9 nm) bilayer, which are used to obtain the field-like torque coefficient. The imaginary component of the spin mixing conductance can then be determined numerically by rearranging Equation (6). From this we obtain $\text{Im}(G^{\uparrow\downarrow}) = 4.49 \times 10^{14}$ Sm$^{-2}$ and hence $G_r = 0.3$.

In this work we concentrated on a single NM/FM interface, modelled by a single spin mixing conductance. It should be noted that there is a possibility for another contribution due to back reflection of spins from the second FM surface, particularly in NM/FM/NM multilayers when the FM layer thickness is comparable to or lower than the spin dephasing length. One approach to modelling the effect of the second surface is to use the concept of transmitted spin mixing conductance [44]. As shown in this reference this results in modified SOT strength, but the model now requires an additional spin mixing conductance. Moreover the presence of an additional NM layer will further enhance the ISTT and investigation of such structures is left for future work. However whilst the exact quantitative dependence of the skyrmion Hall angle on skyrmion diameter is expected to change, qualitatively the conclusions are unaffected, namely the ISTT generated at magnetization gradients, such as skyrmions, in multilayers needs to be taken into consideration in addition to the SOT.

Ref. [19] reported on the current-driven motion of Néel skyrmions with diameters of the order 100nm in a Pt/Co/MgO trilayer, with and without disorder. Micromagnetics simulations were carried out to validate experimental results and the drive dependence of the skyrmion Hall effect was not observed to be accounted for by the FL-SOT. An analytical model is shown for damping coefficients α = 0.43, 0.30 and shows a relatively good agreement to the experimental results. It was suggested that a recreation of the SkHA could be achieved using the description of SOT only with the use of large damping factors, whereas if a smaller damping factor was used the result would be significantly different due to model relation given by Equation (9) when $u_{STT} = 0$. It should be noted that there is disagreement in the literature for damping values for ultrathin Co films, with a range of values reported using different methods [45-52], including time-resolved magneto-optical Kerr effect (TRMOKE), ferromagnetic resonance (FMR), tight-binding modelling (TB), and domain wall (DW) motion. The damping generally increases with decreasing film thickness, which is due to a combination of increased effect of spin-orbit coupling and magnon-electron scattering at interfaces. Moreover, the presence of a heavy-metal layer, such as Pt, further increases the damping values due to spin pumping. Large values of damping have been reported, obtained indirectly from DW motion experiments, namely $\alpha = 0.43$ for Pt(3 nm)/Co(0.9 nm) [19], and $\alpha = 0.3$ for Pt(4.5 nm)/Co(0.8 nm)/Pt(3.5 nm) [52] in the precessional DW motion regime, although when the steady DW motion regime was used values ranging up to $\alpha = 3.4$ were reported [52]. On the other hand, direct measurement techniques have shown much smaller values for comparable systems, typically of the order $\alpha = 0.1$ or below. Thus FMR measurements obtain $\alpha = 0.03$ for Pt(1.5 nm)/Cu(10 nm)/Co(1 nm)/Cu(10 nm) [46], $\alpha = 0.08$ for Pt(1.5 nm)/Co(1 nm)/W(1.5 nm) [46], $\alpha = 0.02$ for Co(1.4 nm)/Pt(1 nm) [45] and $\alpha = 0.18$ for Pt(5 nm)/Co(1 nm)/Pt(2 nm) [51]. Similarly results from TRMOKE measurements obtain values of $\alpha = 0.13$ for [Co(0.4 nm)/Pt (0.8 nm)]$_{12}$ [50], $\alpha = 0.2$ for Pt(5 nm)/Co(1 nm)/Pt(2 nm) [51], and $\alpha = 0.03$ extrapolated for Pt(4 nm)/Co(1 nm)/AlOx. TB modelling on Co(1 nm)/Pt(1.2 nm) also obtains a comparable value of $\alpha = 0.08$ [48].

It is well known that for skyrmions an additional contribution arises due to the emergent magnetic field and emergent time-dependent electric field, resulting in a topological Hall effect (THE) and charge pumping effect respectively [42,53,54]. As has been shown previously this results in an enhancement in the non-adiabaticity of vortex structures and skyrmions [42,53]. In the case of ISTT however the effective non-adiabaticity is strongly dependent on the FM layer thickness and the NM layer spin diffusion length, and for the Co/Pt bilayer considered here $|\beta_\perp| > 1$ as shown in Appendix B. As we have verified using an extended drift-diffusion model which includes the emergent electromagnetic field (for model implementation details see [55]), the effect on the skyrmion motion in the Co/Pt bilayers studies here is negligible, however further investigations are required to characterize the different possible regimes, including combinations of ISTT and bulk STT together with the THE and charge pumping, particularly as a function of FM layer thickness and NM layer transport properties.

# IV. Skyrmion Characteristics on Disordered Surfaces

In-depth theoretical investigations of skyrmions interacting with defects have shown significant changes to the skyrmion paths [18,25,28,56-60] with further investigations showing disorder to have a large effect on the SkHA and skyrmion velocity [14,16,19,28,38]. There is also an inherent problem when using the Thiele equation for comparison due to the inability to incorporate skyrmion malleability. Furthermore, we investigate the effect surface disorder has on the SkHA with the ISTT component included in the simulations. This was accomplished by using a similar method to that used for the disorder-free results, however, in this case, a surface disorder is applied which is accomplished using a jagged surface roughness profile [28]. To investigate the effect of disorder on skyrmion motion, three surface roughness variation periods were applied, with 20 nm, 40 nm and 60 nm average spacing respectively between peaks and troughs. Fig. 1(b) shows a disorder variation period of 60nm. Furthermore, in the ideal case a single applied current was used as the SkHA values are independent of the applied current density, but in the disordered case a range of applied current densities were used, ranging from $J_c = 6$ GAm$^{-2}$ up to $J_c = 625$ GAm$^{-2}$.

Fig. 1(b) shows three simulated skyrmion paths which display three regimes: pinned, creep and flow regime. These have been defined when being driven by current as: (i) the pinned regime in which the skyrmion is confined within a potential well due to the energy landscape surrounding the skyrmion, and displays an orbiting path inside the pinning potential. (ii) The creep regime, in which the skyrmion has a large enough driving force to overcome the energy barrier and unpin itself, however, the skyrmion is heavily affected by the local energy landscape resulting in a significantly distorted path via sporadic jumps between pinning sites. (iii) The flow regime, in which the skyrmion displays a near-linear path across the disorder landscape, close to that obtained in the ideal case.

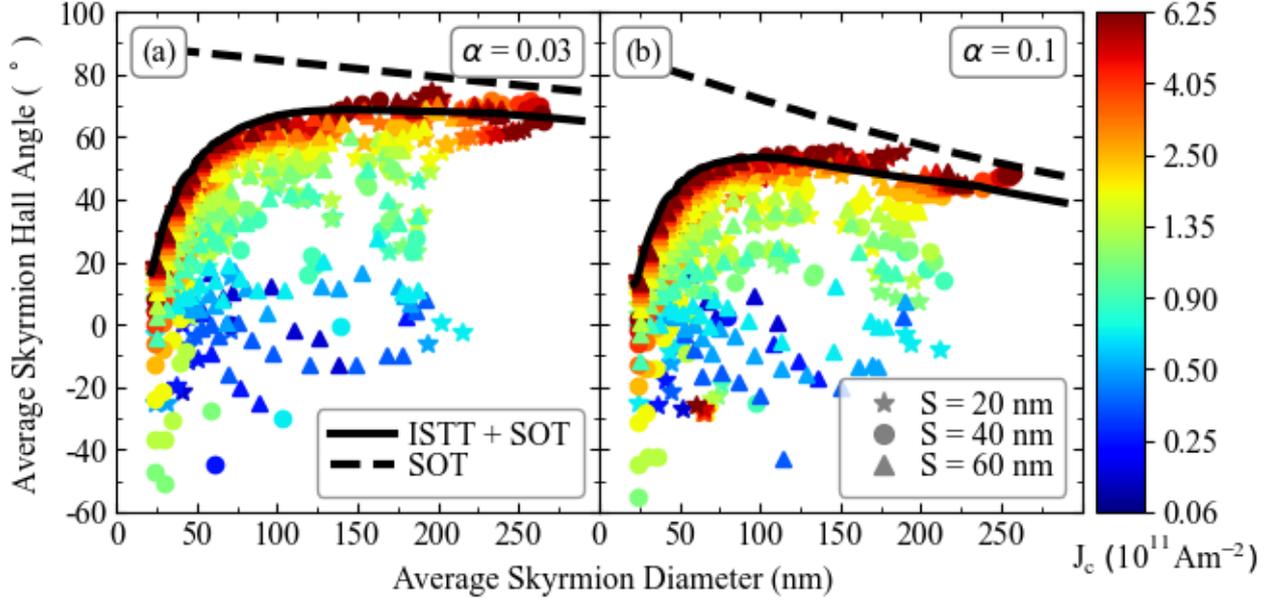

FIG. 3. Skyrmion Hall angle dependence on skyrmion diameter for damping values of (a) $\alpha = 0.03$ and (b) $\alpha = 0.1$, with varying surface disorder periodicity in a Co (1 nm)/Pt (3 nm) bilayer. Star, circle, and triangle correspond to 20, 40 and 60nm disorder period, respectively. The colour map corresponds to the applied current density in the +x direction of the Co/Pt system. The solid black lines display the full spin transport solver results in the ideal case. The dashed black lines show the SOT dependence obtained from the Thiele equation solution with $u_{STT} = 0$.

Fig. 3 shows the relationship between the SkHA and skyrmion diameter for $\alpha = 0.03$ and $\alpha = 0.1$. The solid black line is the trend obtained in the ideal case for the full spin torque, and the dashed black line shows the expected behaviour under SOT using the rigid skyrmion model. The SkHA increases rapidly with skyrmion diameter up to 100 nm, after which a flattening of the trend occurs. The maximum SkHA obtained is $\theta_{SkHA} = 75°$ for $\alpha = 0.03$, and $\theta_{SkHA} = 55°$ for $\alpha = 0.1$. At small driving currents skyrmions experience high deflection due to the local energy landscape, which is typical for the creep regime [61] where skyrmion motion occurs via sporadic jumps between pinning sites. In this regime, the SkHA is nearly independent of the skyrmion diameter, in agreement with experimental observations [14]. At larger driving currents a more defined SkHA behaviour becomes apparent, tending towards the ideal case and entering the flow regime. There are two distinct regions at larger driving currents: (i) skyrmions with diameters of 100 nm and less which show a rapid SkHA increase with increasing diameters due to the combined action of the SOT and ISTT, and (ii) skyrmions with diameters greater than 100 nm that show a saturation of the SkHA, which becomes near-constant. Thus the effect of the combined SOT and ISTT model is in sharp contrast to the Thiele equation solution in Equation (9) with $u_{STT} = 0$, which predicts a monotonic increase in the SkHA with decreasing skyrmion diameter.

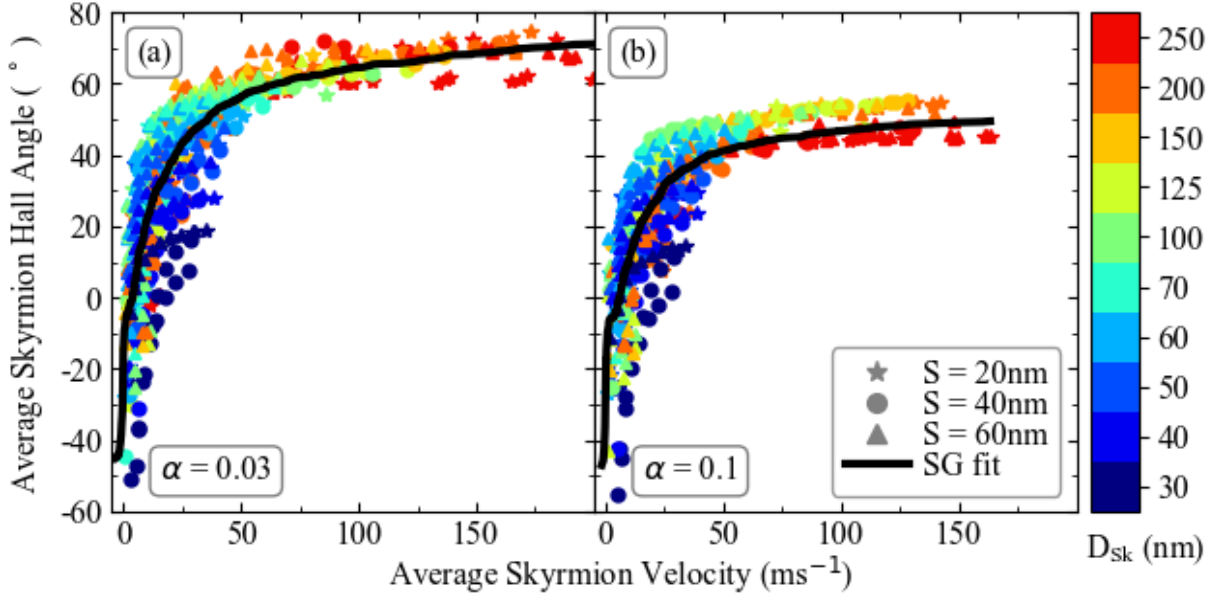

FIG. 4. Skyrmion Hall angle-dependence on the skyrmion velocity for three disorder variation periods. Two damping values in (a) $\alpha = 0.03$ and (b) $\alpha = 0.1$ were used to obtain the dependence. The colour map details the diameter bin ranges from the minimum range of less than 30 nm up to a maximum of 250 nm. The points labelled in (a) and (b) correspond to the disorder variation periods on the Co(1 nm)/Pt(3 nm) bilayer. Star, circle and triangle correspond to 20, 40 and 60 nm spacings, respectively. The solid black line shows the Savitzky-Golay trend line.

Ref. [14] observed skyrmion characteristics experimentally, analyzed using a range of damping factors. Using the Thiele equation with SOT only, significant discrepancies are observed between modelling and experimental results. In particular, the experimental results show a SkHA largely independent of skyrmion diameter, whereas the model based on the Thiele equation predicts a significant increase in SkHA with decreasing skyrmion diameter, and this discrepancy is significantly pronounced below a diameter of 100 nm and has not been reproduced using micromagnetics simulations to date. The results obtained here show the ISTT contribution is particularly important for small skyrmion diameters, which can lead to a near diameter independent SkHA as observed experimentally, particularly when defect energy landscapes are taken into consideration.

Ref. [16] shows interesting skyrmion temperature characteristics in which skyrmion velocity is strongly dependent on temperature due to the increase in the temperature-dependent DL-SOT. However, the temperature is shown to have little effect on the SkHA, based on a model of spin structure deformations due to the FL-SOT, as the relationship between velocity and SkHA collapse onto a universal trend line. This behaviour has also been simulated [28] in which the skyrmion path and SkHA deviate insignificantly at T = 0 K and T = 297 K in a Pt/Co/Ta stack. When considering both ISTT and SOT components in our model the SkHA and skyrmion velocity show similar characteristics to the experimental results, as shown in Fig. 4. This shows the SkHA as a function of the skyrmion velocity, with the skyrmion diameter colour-coded as shown in the legend, for the three different disorder variation periods. The black line shows the Savitzky-Golay fit to the data to illustrate the average trend of the SkHA. There are two distinct velocity behaviours apparent in Fig. 4. Initially, there is a steep increase in the SkHA with an increasing velocity at mainly small skyrmion diameters, indicative of the creep regime, which agrees well with established defect theory [18] and experimental works [14,16,19]. Following this initial steep increase in SkHA, a flattening of the behaviour occurs at which a much more gradual increase is observed over a much larger velocity range. These results for a range of skyrmion diameters suggest a collapse onto a universal SkHA versus skyrmion velocity dependence which is nearly independent of the surface roughness characteristics, similar to experimental results obtained in similar systems [14,16,18,19].

Fig. 5 shows the velocity dependence on the applied current. The relationship for the ideal case is shown by the coloured coded dashed lines. A current-velocity relationship is subsequently established for both α = 0.03 and 0.1, indicating a monotonic increase of the average skyrmion velocity, dependent on the applied current density. The disordered simulations correspond very well to simulations of no disorder in which the effect of surface roughness mainly affects the SkHA by deflecting skyrmions and does not have an appreciable influence on the skyrmion velocity. Furthermore, the damping values have a significant effect on the skyrmion velocity which is predicted by the Thiele equation and is most evident in the 250 nm bin at the largest applied current, where for α = 0.03 and α = 0.1 the skyrmion velocities are $v_{Sk}$ = 191 ms$^{-1}$ and $v_{Sk}$ = 163 ms$^{-1}$ respectively. As pinned skyrmions are not shown, the skyrmion velocity at low driving current displays slight stochasticity for $J_c$ < 1×10$^{11}$ Am$^{-2}$, after which the skyrmions have a well-defined velocity with respect to the applied current. Furthermore, for $J_c$ > 2.5×10$^{11}$ Am$^{-2}$ the skyrmion velocity tends to the linear profile of the disorder-free case. This behaviour corresponds well with experimental behaviour [18,19,59]. A more detailed depiction of the velocity-current-density relationship at small current densities is shown in Appendix C.

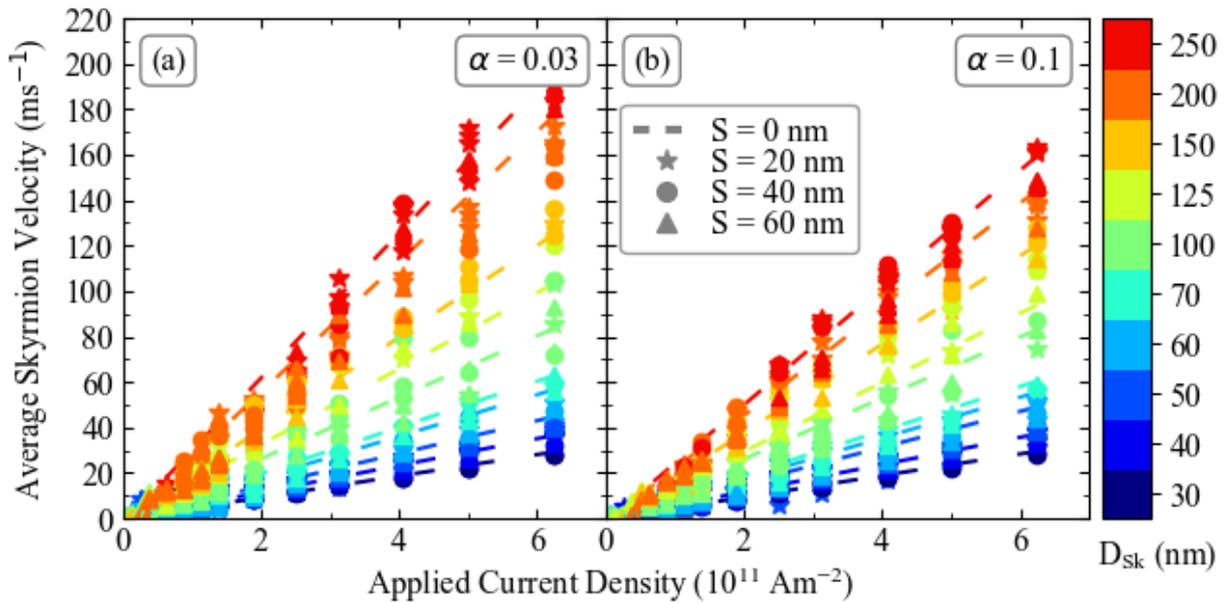

FIG. 5. Relationship between skyrmion velocity and applied current density for damping values of (a) $\alpha = 0.03$ and (b) $\alpha = 0.1$, with surface disorder periodicity in a Co (1 nm)/Pt (3 nm) bilayer. Star, circle, and triangle correspond to 20, 40 and 60 nm disorder period respectively. Each skyrmion was assigned into a bin with assigned diameter ranges. The colour key details the diameter bin ranges from 30 nm up to a maximum of 250 nm. The dashed lines correspond to the relationship for the ideal (S = 0 nm) case at each diameter range.

Fig. 6 shows the relationship between the skyrmion threshold current and skyrmion diameter for $\alpha = 0.03$ and $\alpha = 0.1$. We define the threshold current as the current required to change the skyrmion from the pinned regime to the creep regime which has been referred to as the depinning current. For $\alpha = 0.1$ the required current to unpin the skyrmions is larger than for $\alpha = 0.03$ at small skyrmion diameters, however for larger diameters the threshold currents for both $\alpha = 0.1$ and $\alpha = 0.03$ tend towards each other. Typically for small skyrmions, the threshold current density is significantly greater than for larger skyrmions, which is most likely due to greater skyrmion rigidity at smaller diameters. Thus, as the diameter increases the skyrmions become more malleable and start to deform as diametrically opposite points on the skyrmion border experience different pinning energies. This helps to reduce the threshold current, as the exchange energy built up through skyrmion deformations assists in depinning the skyrmion border for skyrmions much larger than the disorder variation period.

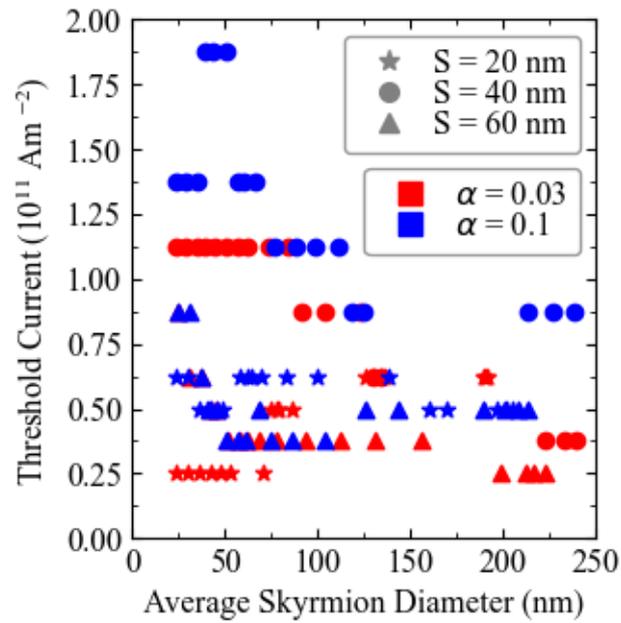

FIG. 6. Skyrmion threshold current for damping values α = 0.03 and α = 0.1 represented by the solid and dashed lines, respectively. Varying applied disorder variation periods of 20, 40 and 60 nm were also used which are represented in red, green, and blue, respectively.

# V. Conclusions

We have studied single skyrmion motion in ultrathin Pt/Co bilayers with and without disorder for a large range of skyrmion diameters utilizing micromagnetics simulations coupled with a self-consistent spin transport solver. As well as SOT acting on the Co layer arising from the SHE, the ISTT was also shown to have a significant contribution to skyrmion motion. The included ISTT term was shown to significantly reduce the discrepancy for the SkHA between experimental results in the literature and SOT-only modelling for skyrmion diameters less than 100 nm. The SOT-only model is shown to be in agreement for large skyrmions, whilst for smaller skyrmions the inclusion of the additional ISTT term results in decreasing SkHA with decreasing skyrmion diameter. The analysis conducted in this work can also be extended to a collection of skyrmions in a sample which will be of interest in future work. The results obtained for a range of skyrmion diameters are shown to collapse onto a universal SkHA-velocity dependence, nearly independent of the surface roughness characteristics. Our results shed light on the current-induced skyrmion dynamics for a large range of diameters and highlight the current shortfalls of using a purely SOT based model in micromagnetics modelling.

# Methods

All simulations were done using Boris Computational Spintronics software [55], version 2.6. Material parameters used in the simulations are summarised in Table I.

Table I – Material parameters used to model the Pt/Co bilayer.

| Parameter | Value | References |
|---|---|---|
| $|D|$ (Co) | 1.5 mJm$^{-2}$ | [12] |
| $M_S$ (Co) | 600 kAm$^{-1}$ | [12] |
| A (Co) | 10 pJm$^{-1}$ | [12] |
| $K_u$ (Co) | 380 kJm$^{-3}$ | [12] |
| α (Co) | 0.03, 0.1 | [46-52] |
| $g_{rel}$ (Co) | 1.3 | [62] |
| σ (Co) | 5 MSm$^{-1}$ | [63] |
| De (Co) | 0.0012 m$^2$s$^{-1}$ | [63] |
| De (Pt) | 0.004 m$^2$s$^{-1}$ | [63] |
| $\lambda_{sf}$ (Co) | 42 nm | [63-64] |
| $\lambda_J$ (Co) | 2 nm | [65] |
| $\lambda_\varphi$ (Co) | 3.2 nm | [66] |
| $\lambda_{sf}$ (Pt) | 1.4 nm | [30] |
| σ (Pt) | 7 MSm$^{-1}$ | [30] |
| Re($G^{\uparrow\downarrow}$) (Pt/Co) | 1.5 PSm$^{-2}$ | [30] |
| Im($G^{\uparrow\downarrow}$) (Pt/Co) | 0.45 PSm$^{-2}$ | * |
| $\theta_{SHA}$ (Pt) | 0.19 | [30] |

* Value is obtained using Ref. [19] in conjunction with Equation (6) which can be solved for Im($G^{\uparrow\downarrow}$).

Computations were done using cell-centred finite difference discretisation. For magnetization dynamics, the computational cell size used was (2 nm, 2nm, 1 nm). For spin transport calculations the computation cell size was (2 nm, 2 nm, 0.5 nm) for both the Co and Pt layers. The LLG equation was evaluated using the RKF45 evaluation which is an adaptive time step method. Initial relaxation was evaluated using the steepest descent method. All computations were completed on the GPU using the CUDA C framework. Roughness profiles were generated using a jagged granular generator algorithm. Equally spaced coefficients at 20, 40 and 60 nm spacings in the x-y plane are randomly generated. The remaining coefficients are obtained using bi-linear interpolation from the randomly generated points.

The Dzyaloshinskii-Moriya interfacial (iDMI) exchange interaction contribution in the xy plane is included as an effective field contribution, which adds to the direct exchange contribution, and is given by:

$$\mathbf{H} = -\frac{2D}{\mu_0 M_S^2}\left(\frac{\partial M_z}{\partial x}, \frac{\partial M_z}{\partial y}, -\frac{\partial M_x}{\partial x} - \frac{\partial M_y}{\partial y}\right)$$

# Appendix A: Skyrmion diameter dependence on the applied magnetic field

To obtain Fig. A1 an isolated skyrmion was set in the centre of a disorder-free Co (1 nm) / Pt (3 nm) bilayer and allowed to relax. Relaxation of the skyrmion is accomplished by evaluating the LLG with the steepest descent method. After the relaxation stage, under a small applied field, was satisfied, an out-of-plane (+z direction) magnetic field was applied and the skyrmion was allowed to relax under the new field. After this, the skyrmion diameter was obtained in the x and y direction independently and averaged.

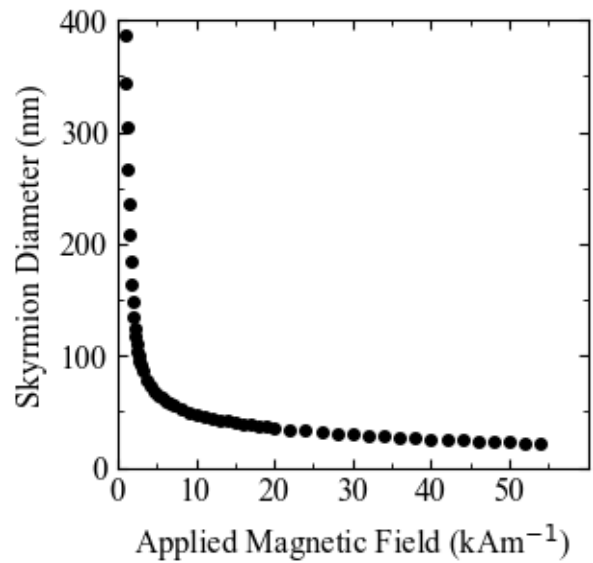

FIG. A1. Skyrmion diameter dependence on out-of-plane magnetic field, computed for the ideal case.

# Appendix B: Dependence of fitting parameters on skyrmion diameters

Fig. A2(a),(b) shows the effective spin Hall angle and field-like SOT fitting parameters for a range of skyrmion diameters, respectively. Fig. A2(c, d) shows the dependence of the effective spin polarization and non-adiabaticity parameters, on skyrmion diameter respectively. There is a slight deviation at small diameters away from the constant value at large diameters. Where dependence is evident, smaller $G_r$ ratios provide a larger effect.

To obtain Fig. A2 a disorder-free Co(1 nm) / Pt(3 nm) bilayer was set using the material parameters stated in Table 1. An isolated skyrmion was then created in the centre of the Co surface, controlled by an out-of-plane magnetic field to tune the skyrmion diameter, and allowed to relax using the steepest descent method. The values for the fitting parameters could then be obtained. Initially, the spin accumulation was computed using the spin transport solver which is used to obtain the total spin-torque. To obtain $\theta_{SHAeff}$ and $r_G$, Equation (4) is fitted to the total spin-torque when negating the ISTT contribution ($P$ set to zero). Similarly, using Equation (7), $P_\perp$ and $\beta_\perp$ can be obtained again as fitting parameters when negating the SOT contribution ($\theta_{SHA}$ set to zero). As we've verified, the full spin-torque computed self-consistently is reproduced using the SOT and ISTT torques with the four fitting parameters, to R values better than 0.99.

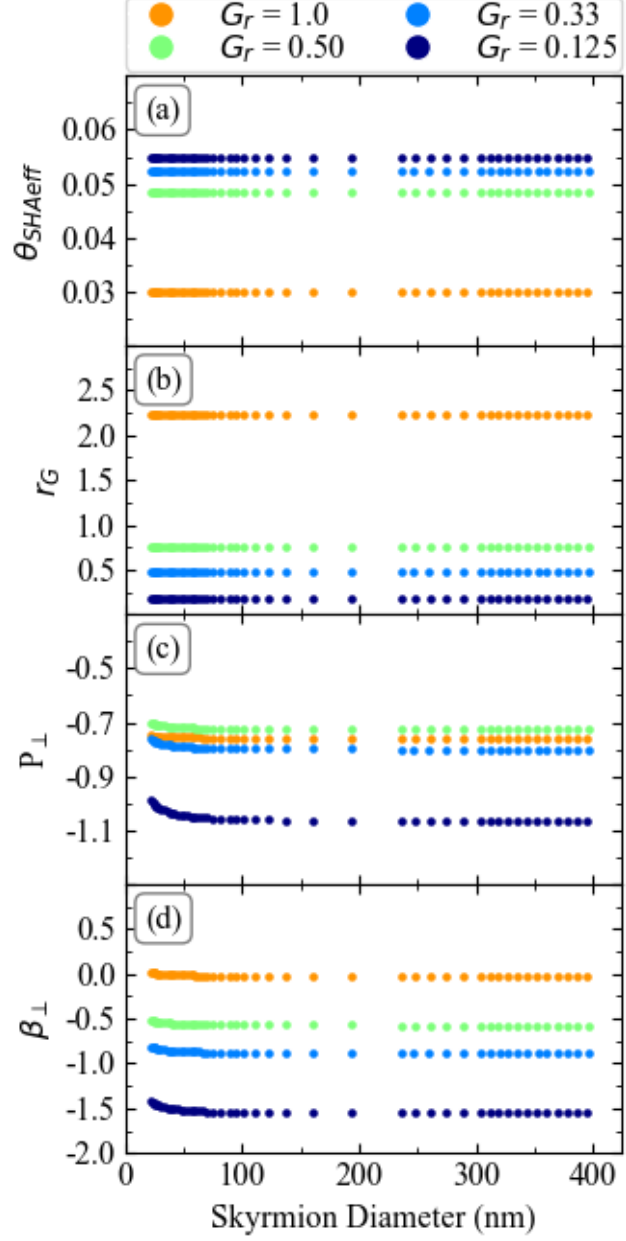

FIG. A2. The dependence of modelling (a) effective Spin Hall angle, (b) field-like spin coefficient, (c) effective spin polarization, and (d) effective non-adiabaticity parameters respectively, on skyrmion diameter in a disorder free Co/Pt bilayer for a range of imaginary to real spin mixing conductance ratios, $G_r$.

# Appendix C: Velocity-current-density relationship at small current densities

Fig. A3 shows the skyrmion velocity for small driving currents which is not visible in Fig. 5 due to the large range. The skyrmion velocity for this range shows a stochastic behaviour, and as a result, do not correspond well with the ideal simulations displayed by the coloured dashed lines.

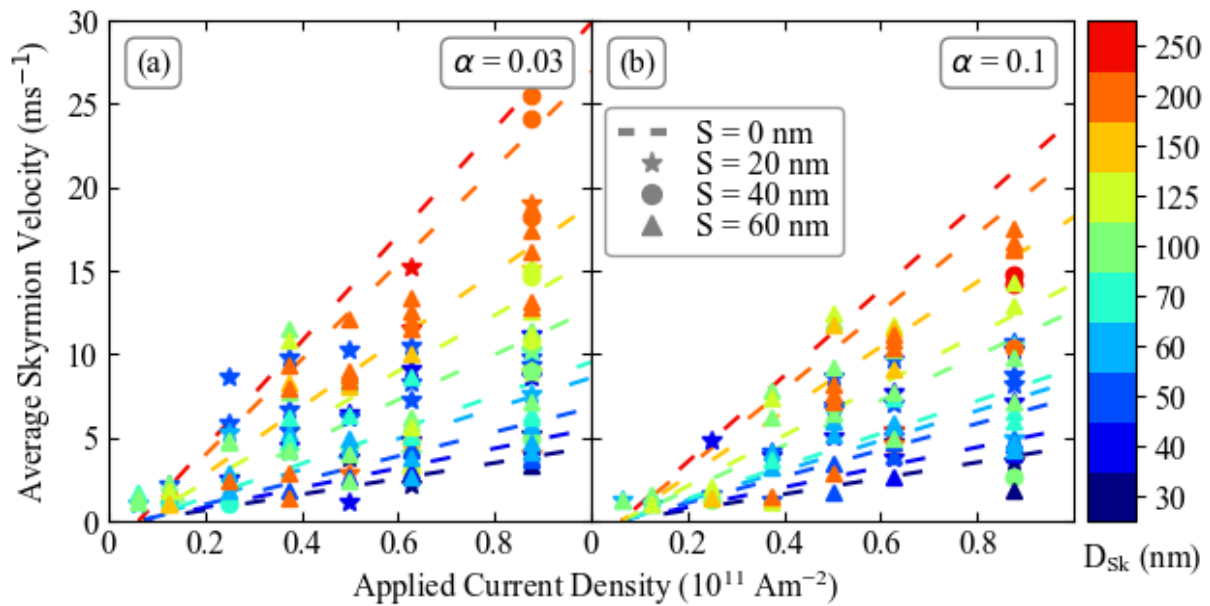

FIG. A3. The velocity dependence on applied current density for (a) $\alpha = 0.03$ and (b) $\alpha = 0.01$. This is a truncated version of Fig. 5 so the stochastic behaviour at lower applied current densities can be seen. The colour key details the skyrmion diameter bin ranges for the minimum of less than 30nm up to a maximum of 250 nm. The dashed lines in (a) and (b) correspond to the relationship for the ideal case (S = 0 nm) at each diameter range.